\documentclass[10pt,conference]{IEEEtran}
\IEEEoverridecommandlockouts

\usepackage{amsmath,amssymb,amsfonts}
\usepackage{algorithmic}
\usepackage{graphicx}
\usepackage{textcomp}
\usepackage{xcolor}

\def\BibTeX{{\rm B\kern-.05em{\sc i\kern-.025em b}\kern-.08em
    T\kern-.1667em\lower.7ex\hbox{E}\kern-.125emX}}

\usepackage[utf8]{inputenc}
\usepackage[
    backend=biber,
    style=ieee,
    citestyle=numeric-comp,
    natbib=true,
    url=false, 
    doi=true,
    eprint=false
]{biblatex}
\usepackage{multirow}
\usepackage{makecell}
\usepackage{ragged2e}
\usepackage{booktabs}
\usepackage{microtype}
\usepackage{tabulary}
\newcolumntype{R}{>{\raggedright\arraybackslash}X}
\usepackage{placeins}
\usepackage[english]{babel}
\usepackage{csquotes}

\begin{document}

\title{Contradicting Motivations in Civic Tech Software Development: Analysis of a Grassroots Project}

\author{\IEEEauthorblockN{Antti Knutas}
\IEEEauthorblockA{\textit{Dept. of Software Engineering} \\
\textit{LUT University}\\
Lappeenranta, Finland \\
antti.knutas at lut.fi}
\and
\IEEEauthorblockN{Dominik Siemon}
\IEEEauthorblockA{\textit{Dept. of Software Engineering} \\
\textit{LUT University}\\
Lahti, Finland \\
dominik.siemon at lut.fi}
\and
\IEEEauthorblockN{Natasha Tylosky}
\IEEEauthorblockA{\textit{Dept. of Software Engineering} \\
\textit{LUT University}\\
Lappeenranta, Finland \\
natasha.tylosky at lut.fi}
\and
\IEEEauthorblockN{Giovanni Maccani}
\IEEEauthorblockA{\textit{Ideas for Change}\\
Barcelona, Spain \\
giovannimaccani at\\ ideasforchange.com}}

\maketitle

\begin{abstract}
Grassroots civic tech, or software for social change, is an emerging practice where people create and then use software to create positive change in their community. In this interpretive case study, we apply Engeström's expanded activity theory as a theoretical lens to analyze motivations, how they relate to for example group goals or development tool supported processes, and what contradictions emerge. Participants agreed on big picture motivations, such as learning new skills or improving the community. The main contradictions occurred inside activity systems on details of implementation or between system motives, instead of big picture motivations. Two most significant contradictions involved planning, and converging on design and technical approaches. These findings demonstrate the value of examining civic tech development processes as evolving activity systems.
\end{abstract}


\section*{General Abstract}
Grassroots civic tech, where people create and use software to create positive change in their community, is based on volunteer and free/libre open source software participants. Earlier survey-based studies indicate that participant motivations are generally in alignment due to participant self-selection. In our case study, we examine one civic tech grassroots software group and their two projects to see how individual motives evolve and contradict as the project progresses. We found that even though participants agree on the big picture, there are significant contradictions in details. Two most significant contradictions involved planning, and converging on design and technical approaches. To address these contradictions and to sustain similar projects, we recommend involving skilled and committed facilitators that can keep volunteers pointed in the same direction and sustain participant engagement.

\section{Introduction}
Grassroots civic tech, or software for social change, is an emerging practice where people create and then use software to create positive change in their community, such as using citizen sensing to record evidence of air pollution to support environmental activism~\cite{knutas2022civic}. It is related to Free/Libre and Open Source Software (FLOSS) movement and participants are diverse, calling themselves for example civic innovation, tech for good, civic crowdsourcing, free software, and community technology~\cite{costanza-chock2018morethancode}. Like their names for the field, the participants' and groups' motivations and working practices are diverse.

Civic tech itself is a wider field~\cite{sifry_civic_2017,saldivar_civic_2019}, with studies for example in: examining the role of information technologies in governance~\cite{borg_digitalization_2018}; supporting participatory democracy practices of technology in society~\cite{holston_engineering_2016}; understanding the role of software in society~\cite{newman_role_2015}; developing agile methods and techniques to support software development in society~\cite{ferrario_software_2014,gama_preliminary_2017} and, more recently architecting smart cities, open data with and for people~\cite{balestrini_city_2017,larrucea_software_2017}. However, civic tech software development processes from a systems perspective have been less often studied, especially when it comes to grounds-up development processes. Similarly, there have been studies on what drives individuals to participate and keep contributing in FLOSS~\cite{barcomb_why_2019,gerosa_shifting_2021} or tech for good~\cite{huang_leaving_2021}, but less analysis on how volunteer motivations interact once the development process is ongoing. We propose that this is an open question that is worth addressing, since development processes in action~\cite{dittrich_what_2016,dittrich2020exploring} evolve based on team needs. A study on motivations of a project in action can reveal how different volunteer motivations can be negotiated towards a compromise.

In this paper, we apply Engeström's expanded activity theory~\cite{engestrom_learning_2015} as a theoretical lens to analyze motivations, how they relate to for example group goals or development tool supported processes, and what contradictions emerge when a civic tech group's development process is in progress. The research case is based on the activities of Code from Ireland civic tech group from 2018 to 2019, with the group structure described in more detail in~\cite{knutas2019software}. During the case study period, Code for Ireland was engaged in two projects: Transparent Water for publishing data on water quality and Finding Vacant Homes for identifying residences that could alleviate the Irish housing crisis.

More specifically, our research question is: \textit{What motivation, motive, or goal contradictions emerge in a grassroots civic tech project in action?}

In the following Section 2, we detail related research on grassroots civic tech and motivations. We continue then in Section 3, about case study and analysis. Findings are presented in Section 4 and conclusions in Section 5.

\section{Related research on motivations in FLOSS and Civic Tech}
Individual motivations for contributors have been studied in free/libre and open source software projects recently by Gerosa et al. \cite{gerosa_shifting_2021}. They found that key motivational factors were social aspects (helping others, teamwork, reputation) and intrinsic motivations, such as fun, learning and altruism. Similarly, Barcomb et al.~\cite{barcomb_why_2019} found that social norms, satisfaction, and community commitment were supporting factors in episodic volunteers. Coelho et al.~\cite{coelho_why_2018} who studied motivations of core developers mainly had similar results in friendly community and opportunity to engage in volunteer work. By contrast, core developers were also motivated by technical quality, such as improving projects they use, lack of project complexity, and lack of complex or buggy code.

Huang et al.~\cite{huang_leaving_2021} studied specifically FLOSS for social good project participant motivations and found that they are more motivated by project impact, such as the societal issue being addressed, and who are the project owners. Community and social characteristics were also important.

Pern and Kitchin~\cite{perng_solutions_2018} analyze frictions from a broader social and cultural systems perspective in civic tech development, using an earlier Irish case study as a data source. They found that complex urban issues can have only partial or imperfect solutions, that can in turn lead to or reveal further complications. The main frictions Pern and Kitchin discuss were stakeholder engagement, shifting goals, and varying commitment from participating developers.

\section{Research context and analysis}
The case we studied is focused on a grassroots group creating civic tech, Code for Ireland, which describes its mission as \textit{``developing innovative and sustainable solutions to real-world problems faced by communities across Ireland, by fostering collaboration with civic-minded individuals, businesses and public sector organizations.''} The research and data collection occurred during 2018 and 2019. The process is summarized in this Section and is reported in further detail in an earlier study on the group~\cite{knutas2019software}.

Case study data includes interviews, development activity logs from GitHub, and an ethnographic study with a researcher embedded with the group for six months in the first half of 2018. Of these, the main data source was semi-structured interviews with six key group contributors. The interviewees included the community leader, a project manager, an open data advocate, and several programmers. The interviews were recorded, transcribed, and analyzed with qualitative data analysis software. The interview guide is archived in the Zenodo repository~\footnote{https://doi.org/10.5281/zenodo.2565434}. Other collected data was used to contextualize the analysis findings.

Other related research on the group has been performed by Perng and Kitchin~\cite{perng_solutions_2018}, who discussed Code for Ireland from a civic and cultural perspective and defined Code for Ireland as a civic hacking group, where ``civic hacking binds together elements of civic innovation and computer hacking, with citizens quickly and collaboratively developing technological solutions''~\cite[pp. 2]{perng_solutions_2018}.

\subsection{Analysis method}
To accomplish our goals, we framed our research as an interpretive case study \cite{walsham_interpretive_1995, runeson_case_2012}, using an iterative qualitative data analysis process aiming towards increasing abstraction~\cite{saldana_coding_2015,miles2018qualitative}. Case study research was found to be the most suitable for the purposes of this research. It is a qualitative approach in which the investigator explores a bounded system (a case in a specific setting/context) over time, through detailed, in-depth data collection involving multiple sources of information, and reports a case description and case-based themes \cite{walsham_interpretive_1995,yin_case_2009,runeson_case_2012}.

Initial codebook for motivations was based on the work of Barcomb et al.~\cite{barcomb_why_2019}, Silva et al.~\cite{silva_theory_2020}, and Gerosa et al.~\cite{gerosa_shifting_2021}. Novel goals and motives were coded inductively when encountered.

For rigour in our research process, we used a selection of techniques by Lincoln and Guba~\cite{lincoln_naturalistic_1985} for qualitative research. We used \textit{prolonged engagement} to get rich data from the context, used \textit{referential adequacy} in the transcription and coding process, and used \textit{peer debriefing} to get additional neutral viewpoints into data analysis.

\subsection{Theoretical lens}
From an activity-theoretical perspective, human life is fundamentally rooted in participation in activites that are oriented towards objects~\cite{sannino_learning_2009}. These activities are driven by their own or collective purposes, mediated by social rules, shared with a community, and facilitated by tools~\cite{engestrom_activity_2000}. For this paper, we use Engeström's cultural-historical activity theory~\cite{engestrom_learning_2015}, which distinguishes short-lived goal-directed actions and long-term activity systems~\cite{engestrom_activity_2000}. This version is suitable for the study of work and technologies~\cite{nardi_studying_1996} and has been used in IS~\cite{dennehy_breaking_2019}, software engineering~\cite{hemetsberger_collective_2009}, and HCI~\cite{clemmensen_making_2016}.

\begin{figure} 
    \centering
    \includegraphics[width=\columnwidth]{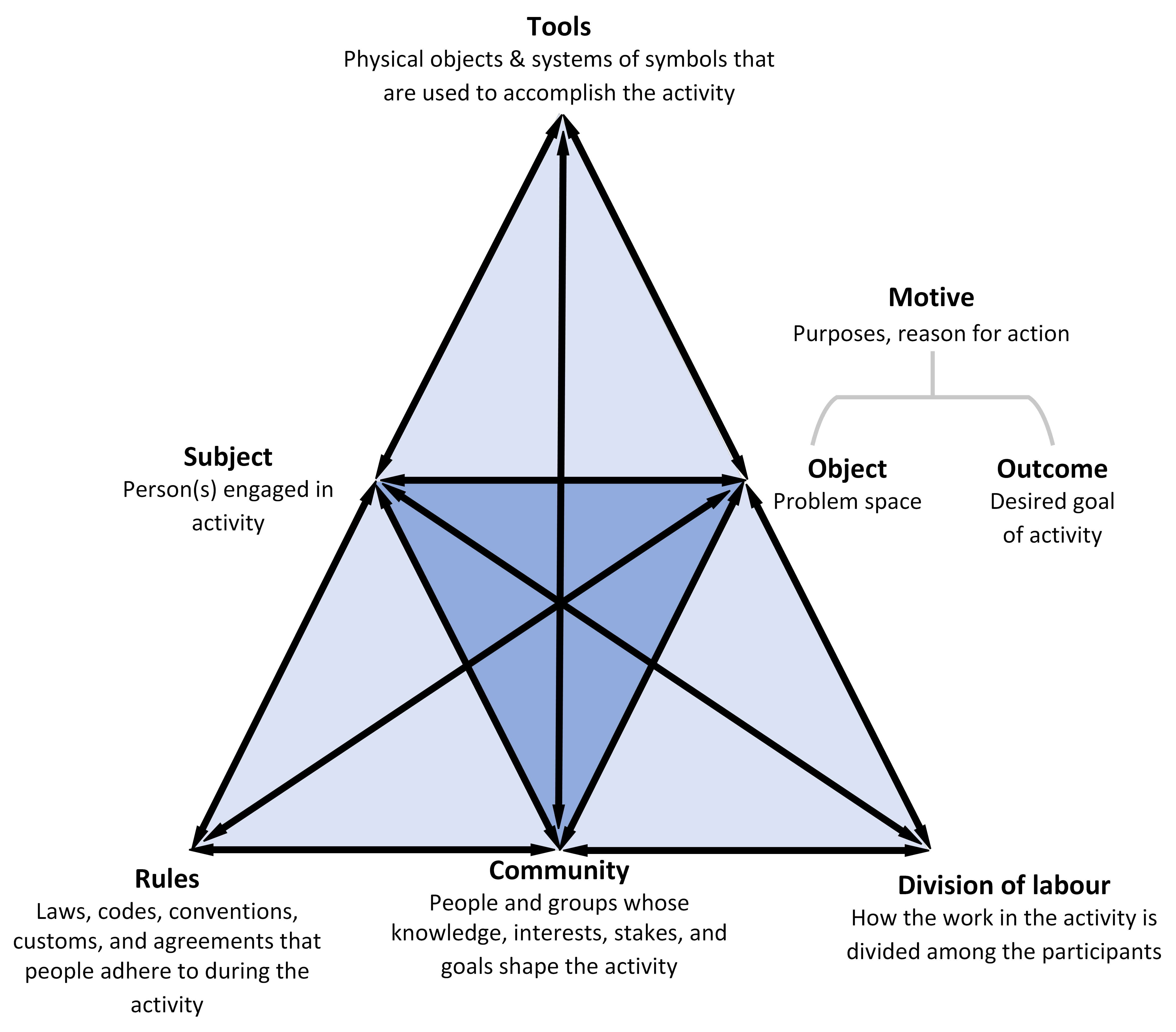}
    \caption{Engeström's activity theory mediation diagram (adapted from~\cite{engestrom_learning_2015})}
    \label{fig:at_triangle}
\end{figure}

In Engeström's cultural-historical activity theory~\cite{engestrom_learning_2015}, activity systems are depicted as a set of mediation triangles, such as in Fig.~\ref{fig:at_triangle}. It displays the subject, object, and facilitating tools. The activity is moderated by rules, community, and the agreed division of labour between actors. Another key concept within activity systems are disturbances and contradictions. While rules, objects and motives give the system stability, they are often contradictory internally or between actors, and also tilt the system towards increasing instability. An activity system therefore constantly evolves towards a new balance based on internal contradictions in what Engeström has named ``cycles of expansive learning.''

Contradictions can occur inside the key concepts or between them. Based on the distances, the contradictions are labeled as follows~\cite{engestrom_learning_2015}: 1) primary (multiple value systems within an activity), 2) secondary (between constituents of the activity), 3) tertiary (between object/motive of the central activity and culturally more advanced new activity), or 4) quarternary (between adjacent activities).

When applying Engeström's cultural-historical activity theory as a theoretical lens, we use the diagram representation to contextualize the motives and contradiction levels to analyze where the contradictions in motives occur. In the following sections where we present the analysis, we use terminology from the theory, which discusses actions, outcomes, and motives, instead of the adjacent word motivation. In this paper, we define motivation as a big picture ``energization and direction of behavior''~\cite{thrash2008approach}. By contrast, motive is a specific reason for taking an action.

\section{Findings}
In this Section, we present our qualitative data analysis outcomes step by step as we progress towards increasing level of abstraction, following lightweight qualitative data analysis practises by Saldana~\cite{saldana_coding_2015} and Miles et al.~\cite{miles2018qualitative}.

In the first step of analysis, open coding, we analyze the interviews and note any motivations and motives. In the second round of analysis, we extract activity descriptions from interviews and online materials, and relate the motives to activities using our theoretical lens. In the final step of analysis, where we generate a situated explanation of the case, we present levels of contradictions inside and between the activities.

\subsection{Discovering Motives in Case}
\label{section:dismoti}
In this Section, we present the findings of open coding analysis where we discovered individual motivations and motives, and then performed a sorting exercise to group the discovered motivations into larger themes. Four most common themes are listed in Table~\ref{tab:motivations}, in a descending order of frequency.

\begin{table}[]
\caption{Most common motivations mentioned by the case participants}
\label{tab:motivations}
\begin{center}
\begin{tabulary}{0.9\columnwidth}{p{0.2\columnwidth}L}
\toprule
\textbf{Motivation} & \textbf{Representative example} \\
\\ \hline
To have an practical impact on society & ``Okay, we're working on software projects. It's very different, it's something good for the community, it's something to give back.'' --P5 \\
Commitment to community & ``But, really, I'll do whatever I can to try and keep the Code For Ireland going on.'' --P3 \\
Learning and developing skills & ``I do at some point maybe see myself managing some people so getting some project management experience, those help as well.'' --P5 \\    
Enjoyment and fun & ``I think the reason for that is that they like the freedom of these projects because if you have worked in a big company like SAP or Facebook, you're pretty much told what to do.'' --P1
\\ \bottomrule
\end{tabulary}
\end{center}
\end{table}

\subsection{Contextualizing Motives in Activities}
In the second step of analysis, we performed another iteration of open coding to detect activity system concepts in two first stages of the software project: planning and design. The stages were originally discovered and defined in~\cite{knutas2019software}.

Then, we used activity theory mediation triangles to contextualize and visualize the constituent interactions in activity systems of 1) project planning and 2) software design. Finally, we analyzed how the stated big picture motivations translate to individual activity system motives.

Planning and design activity systems were selected for presentation in this paper due to them featuring key contradictions. Implementation (prototyping and design) and evaluation stage analyses were omitted from this short paper due to size and scope limitations.

\subsubsection{Planning Activity}
\begin{figure} 
    \centering
    \includegraphics[width=\columnwidth]{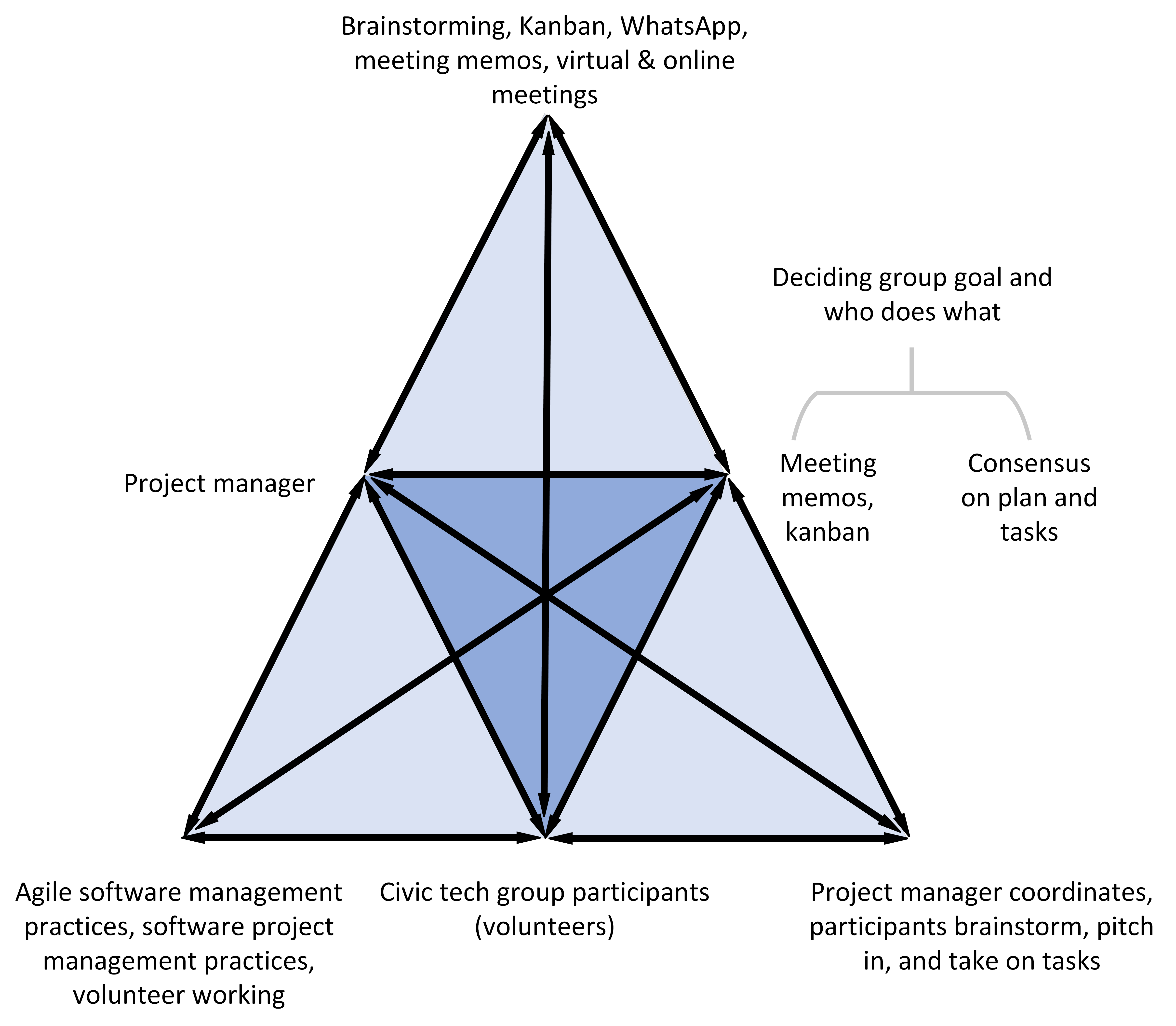}
    \caption{Planning activity system}
    \label{fig:planning}
\end{figure}

In the first activity system, the key motive for project participants is to negotiate software development practises, which current urban issue should be addressed, and work division between current projects. Used tools involve communication tools such Slack and WhatsApp, project management tools such as Trello, and meeting memos. The key actors are the community leader and project managers. The optimal outcome would be a project plan that all participants agree on. The activity system is visualized in Fig.~\ref{fig:planning}.

\subsubsection{Design Activity}
\begin{figure} 
    \centering
    \includegraphics[width=\columnwidth]{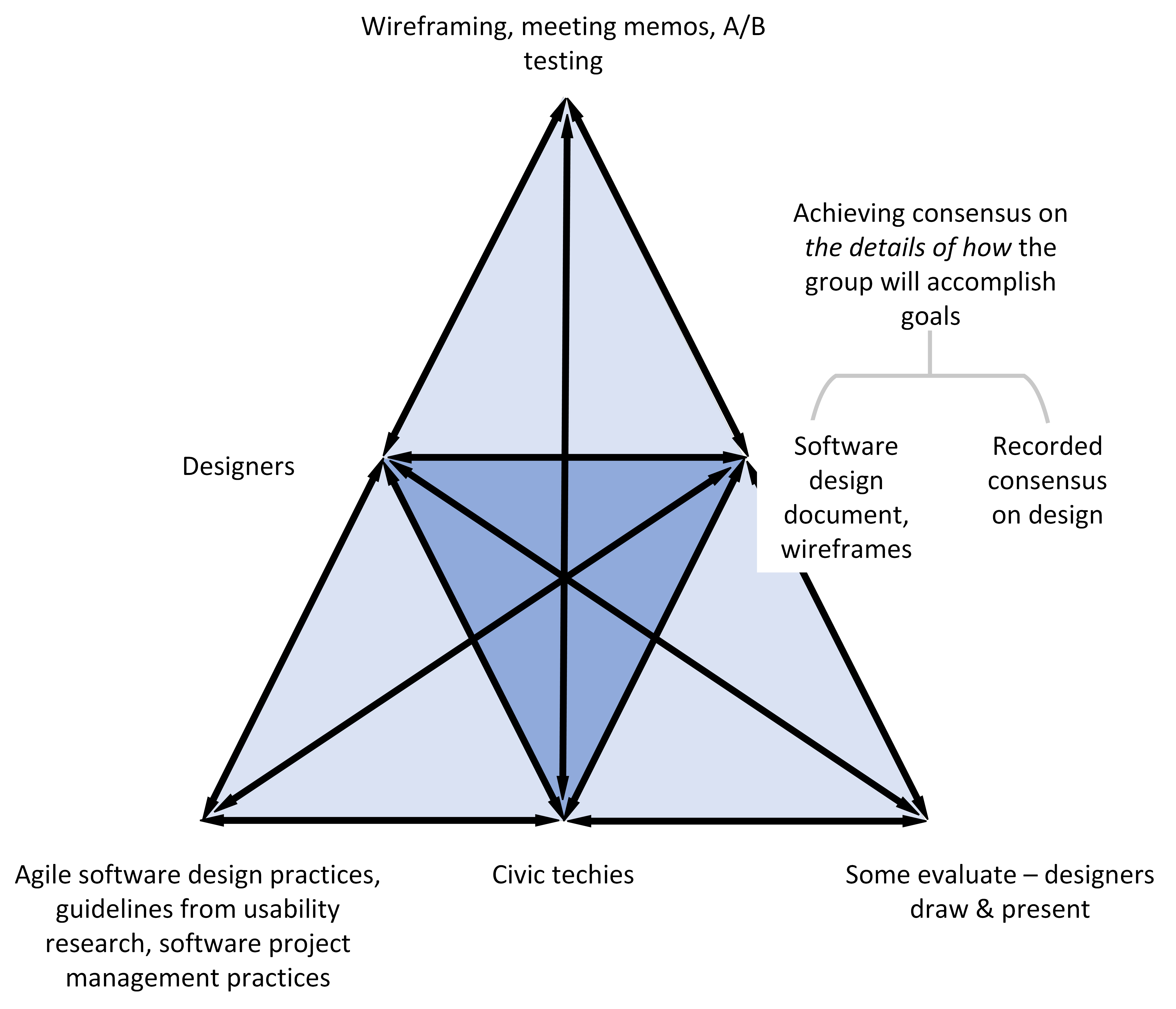}
    \caption{Designing activity system}
    \label{fig:designing}
\end{figure}

In the second activity system, the key motive for a development group participants is to decide in more detail what the final software artefact should be like, and how to divide the work inside the group. Used tools include wire-framing, design sessions, and comparison between visual designs. The outcome of the activity system is a design document. The activity system is visualized in Fig.~\ref{fig:designing}.

\subsection{Contradictions of Motives in Activities}
\begin{table*}[]
\caption{Activity system contradictions}
\label{tab:contradictions}
\begin{center}    
\begin{tabulary}{1.0\textwidth}{LLLL}
\toprule
\textbf{Contradiction level} & \textbf{Activity system(s) and contradicting constituents} & \textbf{Description} \\
\\ \hline
Primary & Design: Rules, tools & There is a difficulty to agree on development tools. Some participants would also like to set strict quality controls and unit tests even at prototype development stage. \\
Primary & Design: Community & Who should be involved in the community during design? (the difficulty in finding opportunities to involve citizens) \\
Secondary & Planning: Community $\Leftrightarrow$ division of labour & Varying level of commitment to agreed on activities between community members. \\
Secondary & Design: Community $\Leftrightarrow$ tools & Contradiction between some participants' skill levels and selected state of the art (but difficult) development tools. \\
Tertiary & Design activity $\Leftrightarrow$ more inclusive design activity & Project participant envisioned better processes, such as co-design or better support in design. However, the current system had no capability to implement them, leading to tensions. \\  
Quarternary & Design activity $\Leftrightarrow$ programming activity & The activity systems were only loosely coupled, programming activity system sometimes disregarding the outputs and motives of design activity.
\\ \bottomrule
\end{tabulary}
\end{center}
\end{table*}

In the final stage of analysis, contradictions inside activity systems, between system concepts, and between activity systems were analyzed. Inside each activity system concept, participant motivations (as listed in Section~\ref{section:dismoti}) were mainly in alignment. Whether participant was motivated by impact, commitment to friends, learning, or fun, they agreed with the overall motives and what the activity system should accomplish. What caused most contradictions inside activity systems was how these goals should be accomplished: in software engineering methods, selection of instruments, and various levels of commitment to agreed on division of labour. Key contradictions identified in analysis and their levels are listed in Table~\ref{tab:contradictions}.

Between activity systems, two main contradictions were loose coupling between software engineering stages (or activity systems).

\subsection{Discussing the Findings}
Some of the largest contradictions emerged in communication and task management, where a phone-based instant messenger was used for communication and task tracking. A second major issue occurred in design and choosing the technical specification: Design went through several rapid iterations, was loosely coupled, and several competing (instead of complementing) technical approaches were simultaneously implemented.

In our case study, we found that the big picture motivations of participants were in alignment and what differs is how participants want to achieve those goals. This follows expectations set by Huang et al.~\cite{huang_leaving_2021}, where they mention that tech for good group membership can be self-selecting for aligned goals.

Our findings indicate that a richer view into participant motives appears when examining a process in action in detail, even though earlier survey-based research indicated that in tech for good projects, motives are initially in alignment. We found that contradictions rose up in the details of decision-making and in varying levels of commitment. These included how to divide tasks, communication tools and details, how to select implementation details, and how to select the technical approach. Majority of the participants were highly motivated, but the volunteer nature of participants and the lack of specialized facilitators meant that good compromises were not achieved on all viewpoints.

\section{Conclusions}
In this paper, we analyzed what motivation, motive, or goal contradictions emerge in a grassroots civic tech project in action. The main contradictions occurred inside activity systems on details of implementation or between system motives, instead of big picture motivations. Two most significant contradictions involved planning, and converging on design and technical approaches.

Our recommendations for practice are what perhaps professionals in established volunteering and non-government organisations already know: Have skilled and committed facilitators that can keep volunteers pointed in the same direction and sustain participant engagement. Based on our case study, it appears that it is essential to have: efficient communicators, shared task management tools, and a compromise on technologies that most participants can commit to. Finding an acceptable lowest common technology denominator is important, since participant skill levels in civic tech can vary~\cite{schrock_what_2019,becker2022smart}.

We extend the body of knowledge in civic tech software engineering through two main contributions. First, we demonstrate activity theory can be used as a theoretical lens to evaluate software engineering methods in action~\cite{dittrich_what_2016,dittrich2020exploring} as evolving, in-action activity systems. Second, we demonstrate that examining participant motives together as an activity system can shed light on contradictions that arise during civic tech or tech for good development.

The trade-off between contextual richness of a single case study and generalizability of a multiple case study points towards a potential limitation in our work. However, seminal literature~\cite{eisenhardt_building_1989,yin_validity_2013} acknowledge single case study as a method of inquiry, especially when undertaking novel, exploratory research to capture the complexity of the context in which the case is situated (i.e. motives and motivations within activity systems in complex civic tech environments and processes). Looking forward, we advocate for more comparisons between different civic tech software engineering activity systems for broader findings and more generalizability.

\section*{Acknowledgment}
We thank Dr. Victoria Palacin for her conceptual contributions and research contributions to the authors' earlier research collaboration. This research is based on the line of thinking that started in collaboration with Dr. Palacin.

\printbibliography 

\end{document}